\documentclass{eas}
\usepackage{graphicx}
\begin{document}

\title{Predicting Stellar Angular Sizes} 
\author{Kaspar von Braun}\address{Max-Planck Institute for Astronomy (MPIA), K\"{o}nigstuhl 17, 69117 Heidelberg, Germany}
\author{Tabetha S. Boyajian}\address{Department of Astronomy, Yale University, New Haven, CT 06511, USA}
\author{Gerard T. van Belle}\address{Lowell Observatory, Flagstaff, AZ 86001, USA}
\begin{abstract}
Our survey of long-baseline infrared and optical interferometry measurements is producing considerable numbers of directly determined stellar angular sizes. We use our sample of 124 high-precision (5\%) angular stellar diameter values and correlate them with stellar magnitude values from the literature to produce empirical relations for main-sequence stars between observed apparent magnitudes, stellar colors, and angular sizes (surface brightness relations). We find a significant dependence on stellar metallicity for ($B-V$) colors. The scatter in the calculated relations is small ($\sim$5\%), which makes them a robust tool for the prediction of main-sequence stellar angular sizes based on photometry. We apply these relations via the calculation of the radius of the multiplanet host star GJ~667C.\end{abstract}
\maketitle
\section{Introduction}
Stellar surface brightness is directly related to stellar broad-band color, and it is also a function of stellar apparent magnitude and angular size (\cite{wes69,bar76}). These dependencies can be combined to show that the stellar angular size can be predicted from stellar photometry, even in the presence of interstellar extinction: 

\begin{equation}
\log \theta_{LD} = -0.2 m_{\lambda}+ \displaystyle\sum\limits_{i=0}^n a_i X^i,
\label{eq1}
\end{equation}

\noindent where $\theta_{LD}$ is the limb-darkening corrected angular stellar diameter in milliarcseconds, $m_{\lambda}$ is the apparent stellar magnitude in the $\lambda$-band, and $X$ is the broad-band stellar color $\lambda_1 - \lambda_2$. The family of functions expressed by Eq. \ref{eq1} for different photometric bands is referred to as ``surface brightness relations" in the literature (see for instance \cite{ker04,bon06,ker08}). 
 
\section{Methods}

In \cite{boy13b}, we fit polynomials of the form of Eq. \ref{eq1} for 48 commonly used color indices in the astronomical literature (e.g., $B-V, V-I$, etc), based on literature broad-band photometry of 124 main sequence stars with interferometrically determined stellar radii with precision of better than 5\%, which were obtained with the CHARA Array\footnote{http://www.chara.gsu.edu/CHARA/} (see \cite{boy12a,boy12b,boy13a}). We provide updates to and refinements of previously published surface brightness relations based on a larger sample of directly measured angular diameters. Typical random errors in the predictions of stellar radii are 4-10\%. We find a metallicity dependence for the $B-V$ colors, but not for the other ones.

\section{Application}

As photometric broad-band magnitudes and colors are (relatively) straightforward to obtain for stars at distances or brightness levels at which interferometric radius measurements are not currently possible, the applications of the surface brightness relations in stellar astronomy are nearly ubiquitous. We use our relations to calculate the stellar diameter of GJ~667C, a late-type dwarf in a triple system, which is known to host between two and seven exoplanets (\cite{ang13}). Application of our surface brightness relations to literature $VJHK_s$ data for GJ~667C, coupled with distance estimates by \cite{del13},  produces a stellar radius estimate of $(0.30 \pm 0.02) R_{\odot}$. 



\end{document}